\documentclass[preprint,12pt]{elsarticle}

\usepackage{amssymb}
\usepackage{amsmath}

\usepackage{hyperref}

\begin{document}

\begin{frontmatter}



\title{QCD in strong magnetic fields: fluctuations of conserved charges and EoS}

\author{Heng-Tong Ding} 
\author{Jin-Biao Gu} 
\author{Arpith Kumar} 
\author{Sheng-Tai Li} 

\affiliation{organization={Key Laboratory of Quark and Lepton Physics (MOE) and Institute of
Particle Physics, Central China Normal University},
            city={ Wuhan},
            postcode={430079},
            state={Hubei},
            country={China}}

\begin{abstract}
Strong magnetic fields can profoundly affect the equilibrium properties, characterized by the equation of state and bulk thermodynamics of strongly interacting matter. Although such fields are expected in off-central heavy-ion collisions, directly measuring their experimental imprints remains extremely challenging. To address this, we propose the baryon-electric charge correlations $\chi^{\rm BQ}_{11}$ and the chemical potential ratio $\mu_{\rm Q}/\mu_{\rm B}$ as magnetic-field-sensitive probes, based on (2+1)-flavor QCD lattice simulations at physical pion masses. Along the transition line, $\chi^{\rm BQ}_{11}$ and $(\mu_{\rm Q}/\mu_{\rm B})_{\rm LO}$ in Pb--Pb collisions increase by factors of 2.1 and 2.4 at $eB \simeq 8M_\pi^2$, respectively. To bridge theoretical predictions and experimental observations, we construct HRG-based proxies and apply systematic kinematic cuts to emulate STAR and ALICE detector acceptances. Furthermore, we extend this investigation to the QCD equation of state, and examine the leading-order thermodynamic coefficients for strangeness-neutral scenarios up to $eB \simeq 0.8 {\rm GeV}^2 \sim 45 m_{\pi}^2$, revealing intriguing non-monotonic structures. 
\end{abstract}


\end{frontmatter}




\section{Introduction}
\label{sec:intro}

Strong magnetic fields---reaching magnitudes comparable to characteristic interaction scales---play an important role in diverse settings from the early Universe and magnetars to relativistic heavy-ion collisions (HIC). They are expected to induce significant non-perturbative effects on the QCD equation of state (EoS) characterizing the thermodynamic properties. One of the most intriguing scenarios occurs in
off-central HICs, where model estimates suggest early-stage strengths $eB\sim5~M_\pi^2$ at the RHIC and $eB\sim70~M_\pi^2$ at the LHC for Pb/Au nuclei collisions~\cite{Deng:2012pc,Skokov:2009qp}.  Such strong fields are expected to alter the hydrodynamic evolution, and can manifest as striking macroscopic phenomena in the produced QCD matter, most prominently the chiral magnetic effect~\cite{Kharzeev:2007jp,Fukushima:2008xe}. The quest to uncover such phenomena has spurred intensive theoretical and experimental investigations, notably aiming to detect imprints in final-state observables.

Both theoretically and experimentally accessible, fluctuations of and correlations among net baryon number (B), electric charge (Q), and strangeness (S) are powerful tools for probing the QCD phase structure~\cite{HotQCD:2012fhj,STAR:2019ans, ALICE:2025mkk} and constructing the QCD EoS~\cite{Bazavov:2017dus}. However, in external magnetic backgrounds, theoretical studies of these fluctuations remain scarce and are largely confined to effective models~\cite{Adhikari:2024bfa}. First-principles lattice QCD calculations are essential to establish model-independent benchmarks. Initial studies in this program employed heavier-than-physical pion mass ($M_\pi\simeq 220~\rm MeV$) at a single lattice spacing~\cite{Ding:2021cwv} and have recently been extended to physical pion mass~\cite{Ding:2023bft,Ding:2025jfz}. Furthermore, these conserved-charge fluctuations have been utilized to construct the QCD EoS in a magnetic background at nonzero density ~\cite{Borsanyi:2023buy,Astrakhantsev:2024mat, Kumar:2025ikm, Ding:2025nyh}.

In this proceedings, we present second-order lattice QCD results for baryon electric charge correlation and electric charge over baryon chemical potential at nonzero magnetic field with physical pions~\cite{Ding:2023bft,Ding:2025jfz}.
To bridge our results with experiment, we employ the Hadron Resonance Gas (HRG) model-based proxies and outline a procedure to implement systematic kinematic cuts to account for detector acceptance limitations. Furthermore, focusing on the HIC scenario, and enforcing strangeness neutrality and isospin asymmetry, we present lattice QCD results for the leading-order EoS pressure coefficient.

\section{Thermodynamics in conserved-charge basis and HRG model }
\label{sec:thm}
The QCD pressure in a magnetized thermal medium, $ P = (T/V)\ln \mathcal{Z} (eB, T, V, {\mu})$, can be expanded as a Taylor series in conserved-charge chemical potentials: 
\begin{eqnarray}
    \hat{P} \equiv \frac{P}{T^4} &=& \sum_{ijk}\frac{1}{i!j!k!}~ \chi^{{\rm B} {\rm Q} {\rm S}}_{ijk}~ \hat{\mu}^{i}_{{\rm B}} \hat{\mu}^{j}_{{\rm Q}} \hat{\mu}^{k}_{{\rm S}}\,, \\
    \chi^{\rm BQS}_{ijk} &=&\frac{1}{VT^3} \left( \frac{\partial}{\partial  \hat{\mu}_{\rm B}} \right)^i \left( \frac{\partial}{\partial  \hat{\mu}_{\rm Q}} \right)^j \left( \frac{\partial}{\partial  \hat{\mu}_{\rm S}} \right)^k \ln \mathcal{Z}\Big|_{\hat{\mu}_{\rm B,Q,S}=0}\,,
    \label{eq:suscp_uds}
\end{eqnarray}
where $\chi^{\rm BQS}_{ijk}$ at leading-order $i+j+k=2$ correspond to fluctuations of and correlations among conserved charges. We compute continuum estimates of these fluctuations and pressure-related observables using lattice QCD simulations on $32^3\times 8$ and $48^3\times 12$ lattices with HISQ action at physical pion mass, focusing around $T_{pc}$ for magnetic strengths up to $0.8~{\rm GeV}^2 \sim 45M_{\pi}^2$. For details on constant U(1) magnetic background, see \cite{Ding:2021cwv,Ding:2025jfz}.

Within the HRG framework, external magnetic fields (along $y$-direction) significantly modify the transverse momentum phase space of charged hadrons, $\int \mathrm{d}^3 \boldsymbol{p} = \int \left|q_R\right| B~ \mathrm{d} l ~\mathrm{d} p_y ~\mathrm{d} \phi_p$. Accordingly, the pressure contribution of individual charged resonance can be written as~\cite{Ding:2021cwv,Ding:2023bft,Ding:2025jfz}: 
\begin{equation}
    \frac{P^c_{R, ~\rm HRG}}{T^4}=\frac{\left|q_R\right| B}{(2 \pi)^3 T^3} \sum_{s_y
    =-s_R}^{s_R} \sum_{l=0}^{\infty} \int_{0}^{\infty}  \mathrm{d} p_y \int_{0}^{2\pi}  \mathrm{d} \phi_p   ~ \sum_{k=1}^{\infty} {(\pm 1)^{k+1}} \frac{e^{-k\left(E^c_R-\mu_R\right)/T}}{k} ,
    \label{eq:pcR}
\end{equation}
for Landau-quantized levels $E^c_R(p_y,l,s_y)=\sqrt{m_R^2+ p_y^2+2\left|q_R\right| B(l+1/2-s_y)}$ with $\mu_R=\mu_{\rm B} {\rm B}_R+ \mu_{\rm Q}{\rm Q}_R+\mu_{\rm S}{\rm S}_R$.

\section{Baryon electric charge correlations as QCD magnetometer}
\label{sec:results_chiBQ}

\begin{figure}[htbp]
\centering
\includegraphics[width=0.49\textwidth,clip]{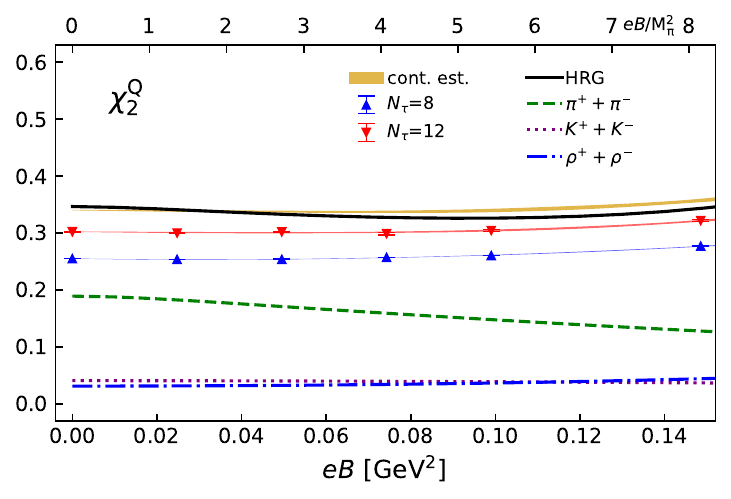}
\includegraphics[width=0.49\textwidth,clip]{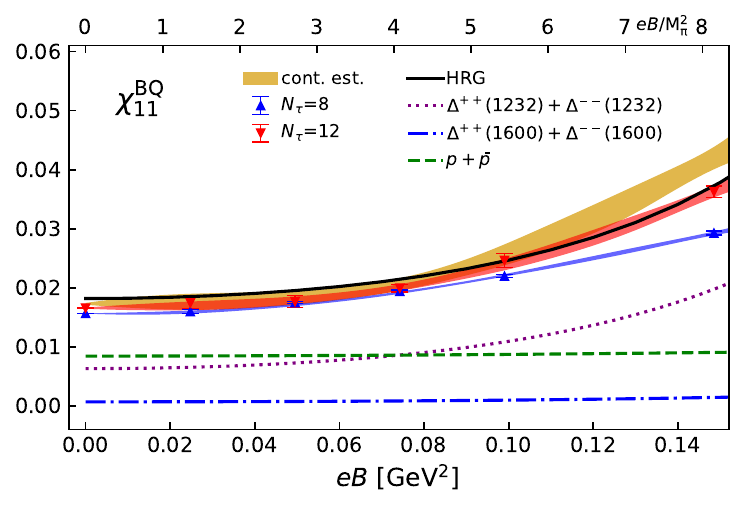}
\caption{Results for $\chi^{\rm Q}_{2}$ and $\chi^{\rm BQ}_{11}$ versus $eB$ at $T=145~{\rm MeV}$. Bands represent lattice data, while solid and broken lines represent HRG results. Figure is taken from Ref.~\cite{Ding:2025jfz}.}
\label{fig:Q2-BQ-T145}       
\end{figure}

Fig.~\ref{fig:Q2-BQ-T145} presents lattice QCD results for $\chi^{\rm Q}_{2}$ and $\chi^{\rm BQ}_{11}$ at $T=145~\rm MeV$. Contrary to naive expectations, the continuum estimate for $\chi^{\rm Q}_{2}$ remains largely unaffected by $eB$, while $\chi^{\rm BQ}_{11}$ exhibits striking sensitivity. From the HRG model, we can discern that magnetic enhancements in $\chi^{\rm BQ}_{11}$ are predominantly driven by the doubly charged $\Delta^{++}(1232)$. However, these $\Delta$ baryons decay rapidly to stable particles, $\Delta^{++}\to p+\pi^+$.

In experiments, these fluctuations are accessed through final-state hadrons: protons ($p$), pions ($\pi$), and kaons ($K$). To align with experiments, we construct proxies using net-conserved charges, i.e., $\text{net-\{B,Q,S\}} \to ~\{\tilde{p},~{Q}^{\rm PID}\equiv \tilde{\pi}^+ + \tilde{p}+\tilde{K}^+,~\tilde{K}^+\}$~\cite{Ding:2023bft,Ding:2025jfz}: 
\begin{equation}
\sigma_{p,Q^{\rm PID},K}^{i,j,k}=\sum_R (\omega_{R\rightarrow \tilde{p}})^i~ (\omega_{R\rightarrow {Q}^{\rm PID}})^j ~\left(\omega_{R\rightarrow \tilde{K}}\right)^k \times I_2^{R}, 
\label{eq:proxies}
\end{equation}
where $I_2^{R} = {\partial^2 (P_R / T^4)}/{\partial \hat{\mu}_R^2}~\big|_{\hat{\mu}_{\mathrm{B}, \mathrm{Q}, \mathrm{S}}=0}$. To reflect detector acceptances, we restrict the momentum-space integration by a Heaviside step function $\Theta$ that enforces kinematic cuts on transverse momentum and pseudo-rapidity,  $\left( \int \mathrm{d} p_y \int \mathrm{d} \phi_p  \right)\times ~ \Theta(p_{T_{\min}}, p_{T_{\max}}, \eta_{\min}, \eta_{\max})$. We then define weights for these cuts as $ \omega^{cuts}_{\pi,K,p}  = I_2^{R\in\{{\pi,K,p} \},~cuts}/I_2^{R\in\{{\pi,K,p} \}} $, which enter Eq.~\ref{eq:proxies} in the following manner: $\sigma_{p,Q^{\rm PID},K}^{i,j,k} :\omega_{R\rightarrow~\tilde{p},\tilde{\pi},\tilde{K}}~~~ \longrightarrow ~~~ \sigma_{p,Q^{\rm PID},K}^{i,j,k;~ cuts} :\omega_{R\rightarrow~\tilde{p},\tilde{\pi},\tilde{K}}~\omega^{ cuts}_{p,\pi,K} $~\cite{Ding:2025jfz,Ding:2025dvn}.

\begin{figure}[htbp]
\centering
\includegraphics[width=0.49\textwidth,clip]{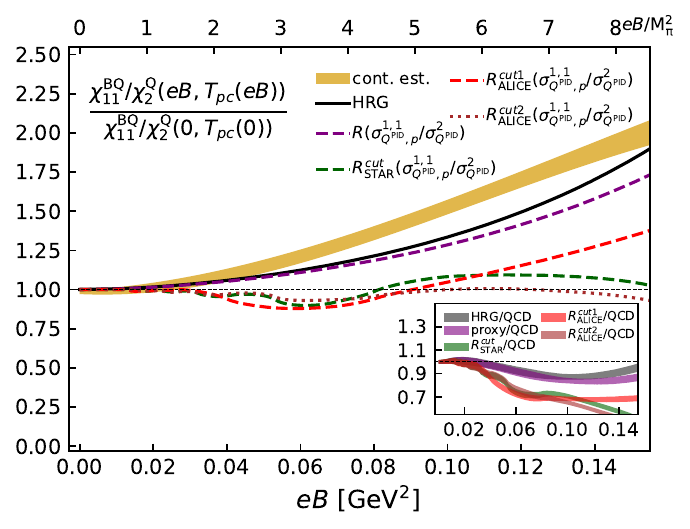}
\includegraphics[width=0.49\textwidth,clip]{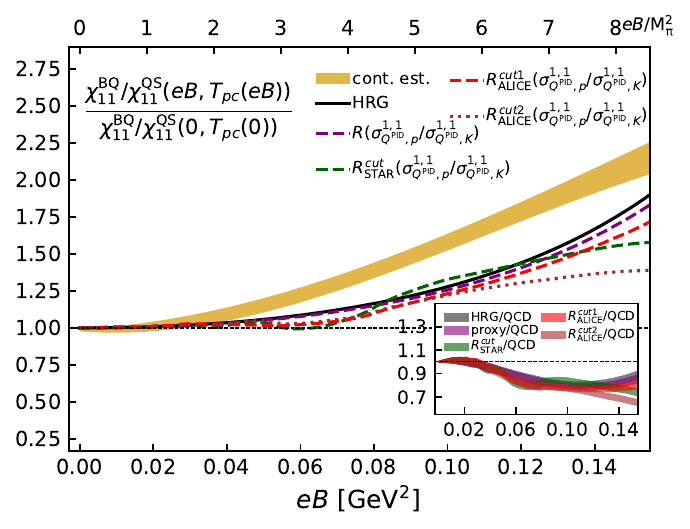}
\caption{$R_{cp}$-like ratio $R\left(\chi^{\rm BQ}_{11} /\chi^{\rm Q}_{2}\right)$ (left) and $R\left(\chi^{\rm BQ}_{11} /\chi^{\rm QS}_{11}\right)$ (right) along transition line $T_{pc}(eB)$. Figure is taken from Ref.~\cite{Ding:2025jfz}.}
\label{fig:ratio_BQ-Tpc}       
\end{figure}

In HICs, the magnetic-field effects are expected to vary across centrality classes. To this extent, we propose $\chi^{\rm BQ}_{11}$-based $R_{cp}$-like (central-peripheral) observables, $R(\mathcal{O}) \equiv \mathcal{O}\left( eB,T_{pc}(eB)\right)~\big/ ~\mathcal{O}\left(eB=0,T_{pc}(0) \right)$, shown for $\mathcal{O} \in \{\chi^{\rm BQ}_{11} /\chi^{\rm Q}_{2},~~\chi^{\rm BQ}_{11} /\chi^{\rm QS}_{11}\}$ in Fig.~\ref{fig:ratio_BQ-Tpc}. Lattice continuum estimates exhibit pronounced enhancements of $\sim2$ for $\chi^{\rm BQ}_{11} /\chi^{\rm Q}_{2}$ (left) and an even more pronounced $\sim2.25$ for $\chi^{\rm BQ}_{11} /\chi^{\rm QS}_{11}$ (right) at $eB\simeq8~{M_{\pi}^2}$. Such remarkable enhancements underscore the potential of $\chi^{\rm BQ}_{11}$ as a magnetometer in QCD. For experiments, these double-ratios are well-suited observables, highlighting $eB$-induced enhancements and suppressing volume-dependent effects~\cite{STAR:2019ans,ALICE:2025mkk}.

Within the HRG framework, for experimental feasibility, Fig.~\ref{fig:ratio_BQ-Tpc} also presents results for corresponding proxies $R( \sigma_{Q^{\rm PID},p}^{1,1} ~\big/~\sigma_{Q^{\rm PID}}^{2}  )$ (left) and $R( \sigma_{Q^{\rm PID},p}^{1,1} ~\big/~\sigma_{Q^{\rm PID},K}^{1,1}  )$ (right), and together with
kinematic cut results, $R^{cut}_{\rm ALICE/STAR}$, emulating STAR/ALICE detector acceptances. As highlighted in the inset, the proxies retain at least $\sim80\%$ of the magnetic lattice QCD sensitivity. Furthermore, incorporating kinematic cuts into proxies still yields increase up to 25\% at $eB\simeq8~{M_{\pi}^2}$ for $\chi^{\rm BQ}_{11} /\chi^{\rm Q}_{2}$, while strikingly up to $60\%$ for $\chi^{\rm BQ}_{11} /\chi^{\rm QS}_{11}$. This underscores the utility of HRG-based proxies to probe magnetic-field signatures in HICs and for bridging theoretical predictions with detector-level analyses. These predictions are undergoing experimental tests~\cite{ALICE:2025mkk,Nonaka:2023xkg}. Alongside ongoing efforts, the ALICE collaboration has already reported centrality-dependent enhancements in the double ratio $\chi^{\rm BQ}_{11}/\chi^{\rm Q}_{2}$~\cite{ALICE:2025mkk}, in qualitative agreement with our theoretical results. We further propose $\chi^{\rm BQ}_{11}/\chi^{\rm QS}_{11}$ as a more sensitive experimental observable.

\begin{figure}[htbp]
\centering
\includegraphics[width=0.49\textwidth,clip]{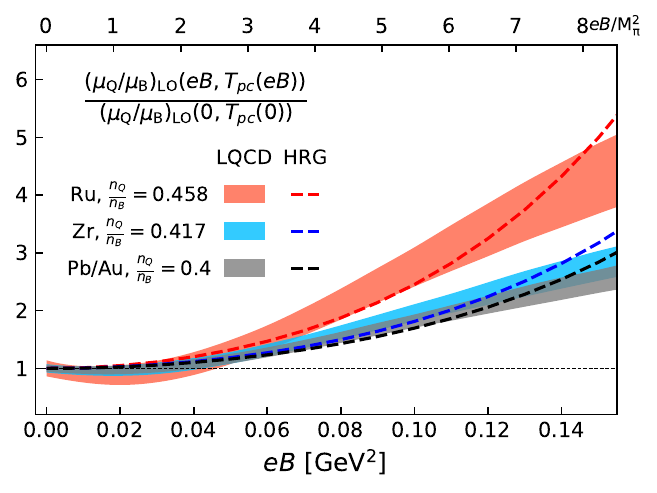}
\caption{$R_{cp}$-like ratio $R(\left(\mu_{\rm Q}/\mu_{\rm B} \right)_{\rm LO})$ along transition line $T_{pc}(eB)$ for various HIC systems. Figure is taken from Ref.~\cite{Ding:2023bft}.}
\label{fig-muQ-muB}       
\end{figure}

Fig.\ref{fig-muQ-muB} presents lattice results for $R_{cp}$-like ratio of electric charge to baryon chemical potential, $R(\left(\mu_{\rm Q}/\mu_{\rm B} \right)_{\rm LO})$, calculated along transition line for various isospin parameters, $n^{\rm Q}/n^{\rm B}$, corresponding to distinct collision systems. In Pb/Au collisions, $R(\left(\mu_{\rm Q}/\mu_{\rm B} \right)_{\rm LO})$ attains approximately 2.4 at $eB\simeq8~{M_{\pi}^2}$ and almost comparable enhancement is observed for the Zr isobar collision system. By contrast, the slightly more isospin-symmetric Ru system exhibits a much steeper rise, reaching $\sim4$ at  $eB\simeq8~{M_{\pi}^2}$, i.e., approximately $1.5$ times stronger magnetic sensitivity than Pb/Au/Zr. Furthermore, HRG model results (broken lines) show good agreement with lattice QCD data. This agreement supports extracting $eB$-dependence of $\mu_{\rm Q}/\mu_{\rm B}$ by fitting particle yields within an HRG framework incorporating magnetized hadron spectrum.

\section{QCD EoS in magnetic fields at nonzero baryon density }
QCD EoS characterizes equilibrium properties and thermodynamic responses of strongly interacting matter under varying control parameters, ($T,eB,\hat{\mu}_{\rm B,Q,S}$). These chemical potentials can be considered as interrelated:
\begin{equation}
 \hat{\mu}_{\rm Q}\equiv q_1 (T,eB) \hat{\mu}_{\rm B} + \mathcal{O}(\hat{\mu}_{\rm B}^3),\quad  \hat{\mu}_{\rm S}\equiv s_1 (T,eB) \hat{\mu}_{\rm B} + \mathcal{O}(\hat{\mu}_{\rm B}^3),
\end{equation}
thereby enabling scenario-dependent simplifications of the QCD EoS.

\subsection{Strangeness neutrality and isospin asymmetry}
\label{subsec:muQovermuB}

In HIC experiments, the colliding nuclei are initially net-strangeness neutral, and their valence quark content constrains the conserved charged densities: $\hat{n}^{{\rm S}} = 0,\quad  {n}^{{\rm Q}} /  {n}^{{\rm B}} = r.$ For $^{208}_{82}$Pb/$^{197}_{879}$Au nuclei, we expect isospin parameter $r \simeq 0.4$, implying slight isospin asymmetry.

\begin{figure}[htbp]
\centering

\includegraphics[width=0.4\textwidth]{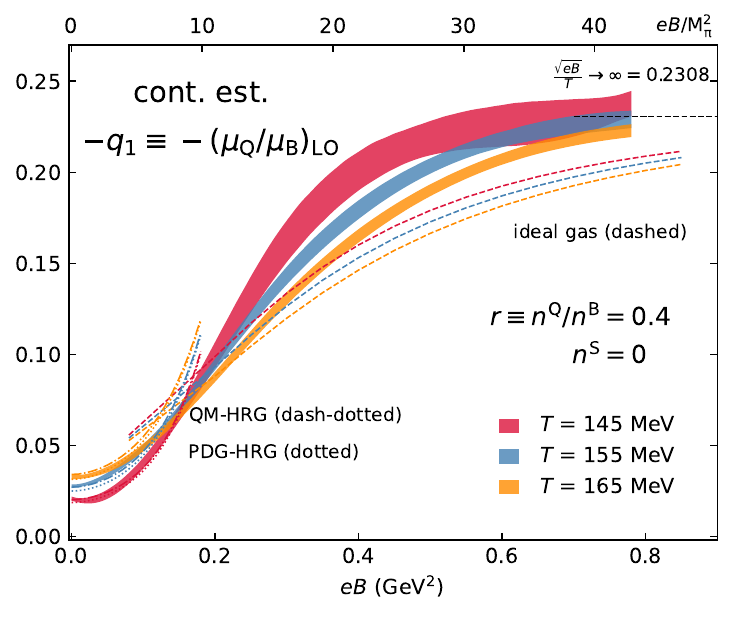}
\includegraphics[width=0.4\textwidth]{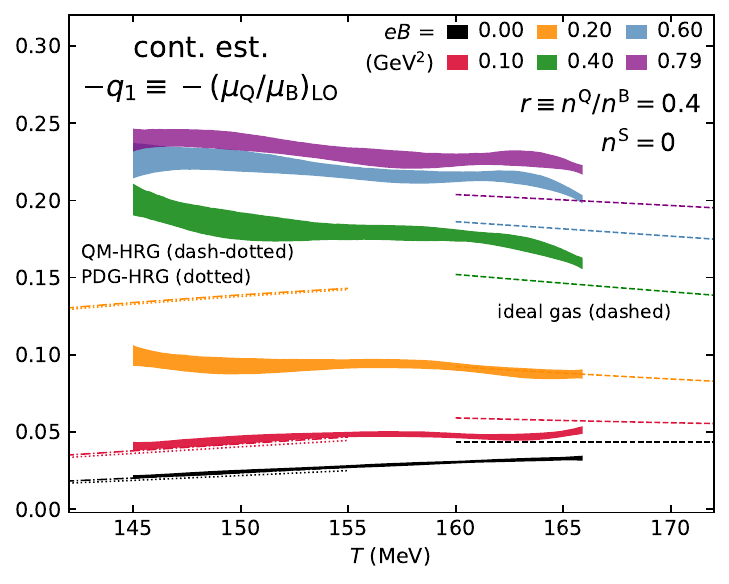}

\caption{Electric charge over baryon chemical potential, $-q_1\equiv -\left(\mu_{\rm Q}/\mu_{\rm B}\right)_{\rm LO}$, for system with $\hat{n}_{\rm S}=0$ and $r\equiv n^{\rm Q}/n^{\rm B}=0.4$. Figure is taken from Ref.\cite{Ding:2025nyh}.}  
\label{fig:q1}
\end{figure}

Fig. \ref{fig:q1} presents lattice continuum estimates for leading--order coefficient
\begin{equation}
    q_1(T, eB) \equiv \left( \frac{{{{\mu}_{\rm Q}}}}{{{{\mu}_{\rm B}}}} \right)_{\rm LO} = \frac{0.4 \left( \chi_{2}^{{\rm B}}  \chi_{2}^{{\rm S}} - \chi_{11}^{{\rm B} {\rm S}} \chi_{11}^{{\rm B} {\rm S}} \right) - \left(\chi_{11}^{{\rm B} {\rm Q}} \chi_{2}^{{\rm S}} - \chi_{11}^{{\rm B} {\rm S}} \chi_{11}^{{\rm Q} {\rm S}} \right)}{\left(\chi_{2}^{{\rm Q} } \chi_{2}^{{\rm S}} - \chi_{11}^{{\rm Q} {\rm S}}  \chi_{11}^{{\rm Q} {\rm S}}\right) - 0.4\left( \chi_{11}^{{\rm B} {\rm Q}} \chi_{2}^{{\rm S}} -\chi_{11}^{{\rm B} {\rm S}} \chi_{11}^{{\rm Q} {\rm S}} \right)}.
\end{equation}
The left and right panels illustrate the $eB$-- and $T$-- dependence, respectively. Lattice results reveal that $q_1$ is negative throughout the $T$–$eB$ parameter space, due to the imposed isospin asymmetry constraint characteristic, where the density of charged baryons is suppressed relative to neutral counterparts. Introduction of magnetic fields further intensifies the negativity of $q_1$.  In the left panel, at magnetic fields around $eB \sim 0.15~{\rm GeV}^2$, crossings among fixed-temperature continuum bands become evident, signaling a reversal of the monotonic temperature hierarchy observed at vanishing and weak magnetic fields. HRG framework fails to capture these crossings, underscoring limits in accounting for intrinsic hadronic structures and non-perturbative QCD effects. In the right panel, this hierarchy reversal manifests as a sign flip in $T$--slope of $q_1$. These phenomena demonstrate a nontrivial interplay between thermal and magnetic effects. Furthermore, lattice results progressively approach saturation in the left panel, converging toward the magnetized ideal gas high-temperature limit, $-q_1(\sqrt{eB}/T \to \infty) = 0.2308$. Note that this saturation arises due to cancellation of the leading linear $eB$ dependence from the dominant lowest Landau level in the fluctuations ratio defining $q_1$~\cite{Ding:2025nyh}. 

\begin{figure}[htbp]
\centering

\includegraphics[width=0.4\textwidth]{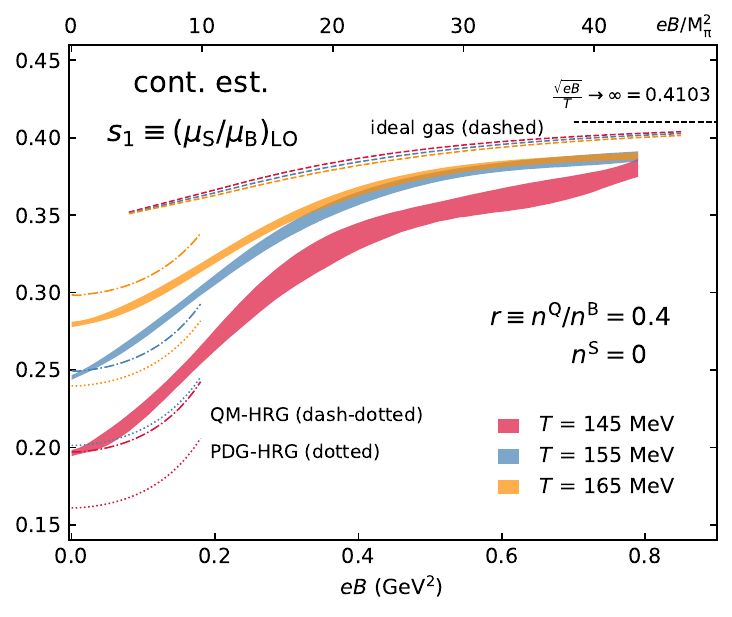}
\includegraphics[width=0.4\textwidth]{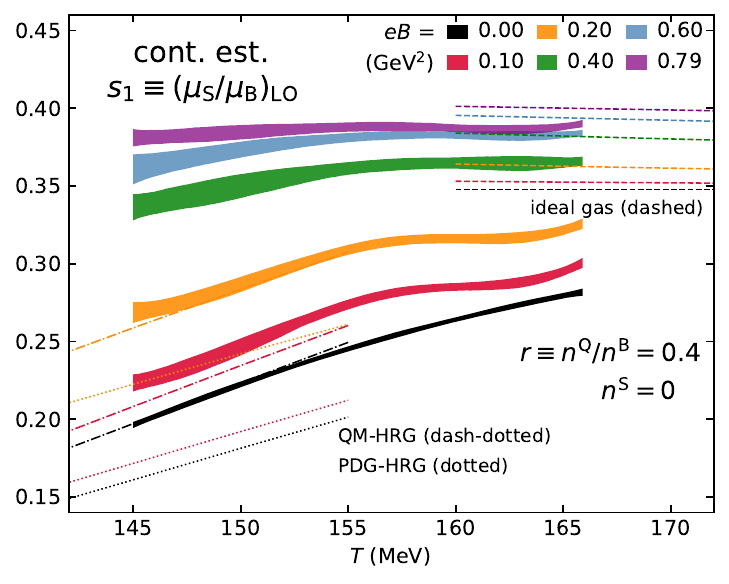}

\caption{Same as Fig.~\ref{fig:q1} but for strangeness over baryon chemical potential, $s_1\equiv\left({{{\mu}_{\rm S}}}/{{{\mu}_{\rm B}}}\right)_{\text{LO}}$. Figure is taken from Ref.~\cite{Ding:2025nyh}.}  
\label{fig:s1}
\end{figure}

The effects of the strangeness neutrality condition are directly encoded in the leading-order coefficient 
\begin{equation}
    s_1(T, eB) \equiv \left( {{{{\mu}_{\rm S}}}}\big/{{{{\mu}_{\rm B}}}} \right)_{\rm LO} = - {\left( \chi_{11}^{{\rm B} {\rm S}} + q_1 \chi_{11}^{{\rm Q} {\rm S}}\right)}/ { \chi_{2}^{{\rm S}}},
\label{eq:s1_def}
\end{equation}
shown in Fig.~\ref{fig:s1}. These effects primarily arise through modifications in baryon-strangeness correlations and the electric charge constraint. Upon introduction of magnetic fields, $s_1$ increases, reflecting similar underlying physics to the enhancement of $-q_1$, but with a positive sign due to the negative strangeness associated with strange quarks. Stronger magnetic fields elevate the charged strange baryons population; strangeness neutrality then demands higher ${{\hat{\mu}_{\rm S}}}$, yielding an increase in $s_1$. The discrepancy between lattice and PDG-HRG results for $s_1$ highlights substantial thermal contributions from additional strange resonances, which are incorporated in QM-HRG. In contrast to $q_1$, no crossings between fixed-temperature bands are observed in $s_1$; instead, the left panel shows a gradual merging of these bands, while the right panel shows a consistent reduction in the $T$--slope for fixed $eB$--bands. This contrast underscores how magnetic fields couple differently to quark mass scales and observable-dependent reorganizations of QCD matter. Analogous to $q_1$, in the magnetized ideal gas limit, $\sqrt{eB}/T \to \infty$ for $T\to \infty$, ratio-observable $s_1$ approaches saturation to $0.4103$~\cite{Ding:2025nyh}.

\subsection{Leading-order pressure coefficient}
\label{subsec:pressure}

\begin{figure}[t]
\centering

\includegraphics[width=0.4\textwidth]{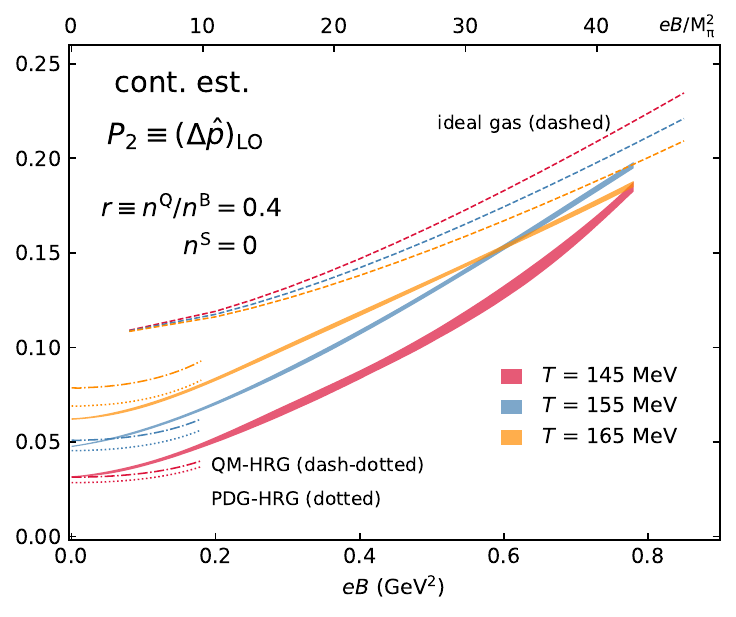}
\includegraphics[width=0.4\textwidth]{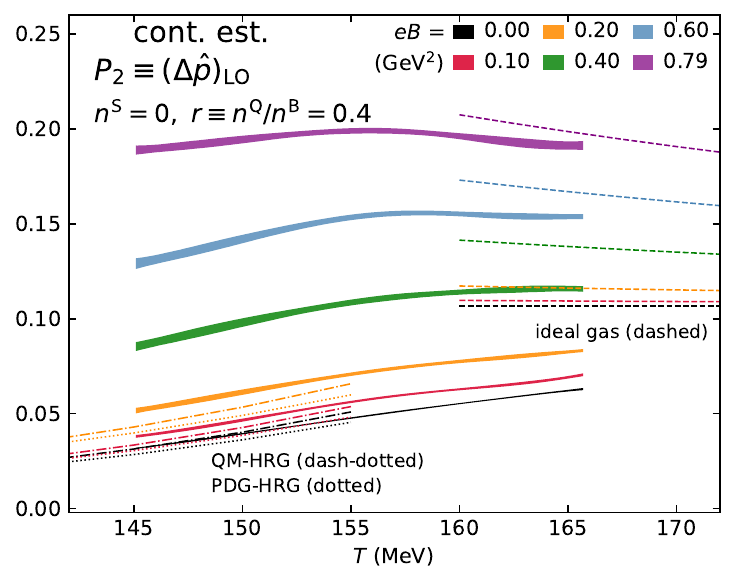}

\includegraphics[width=0.24\textwidth]{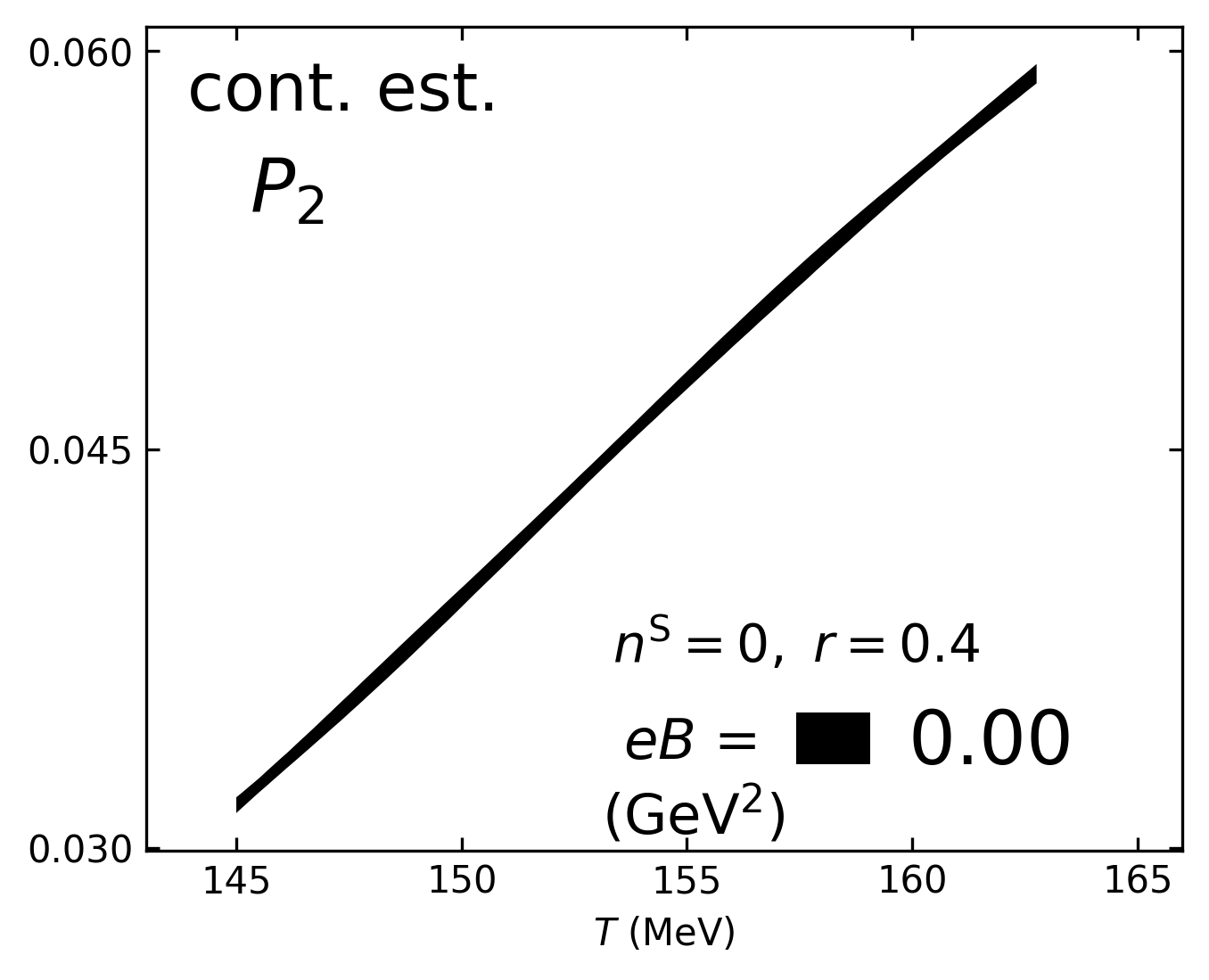}
\includegraphics[width=0.24\textwidth]{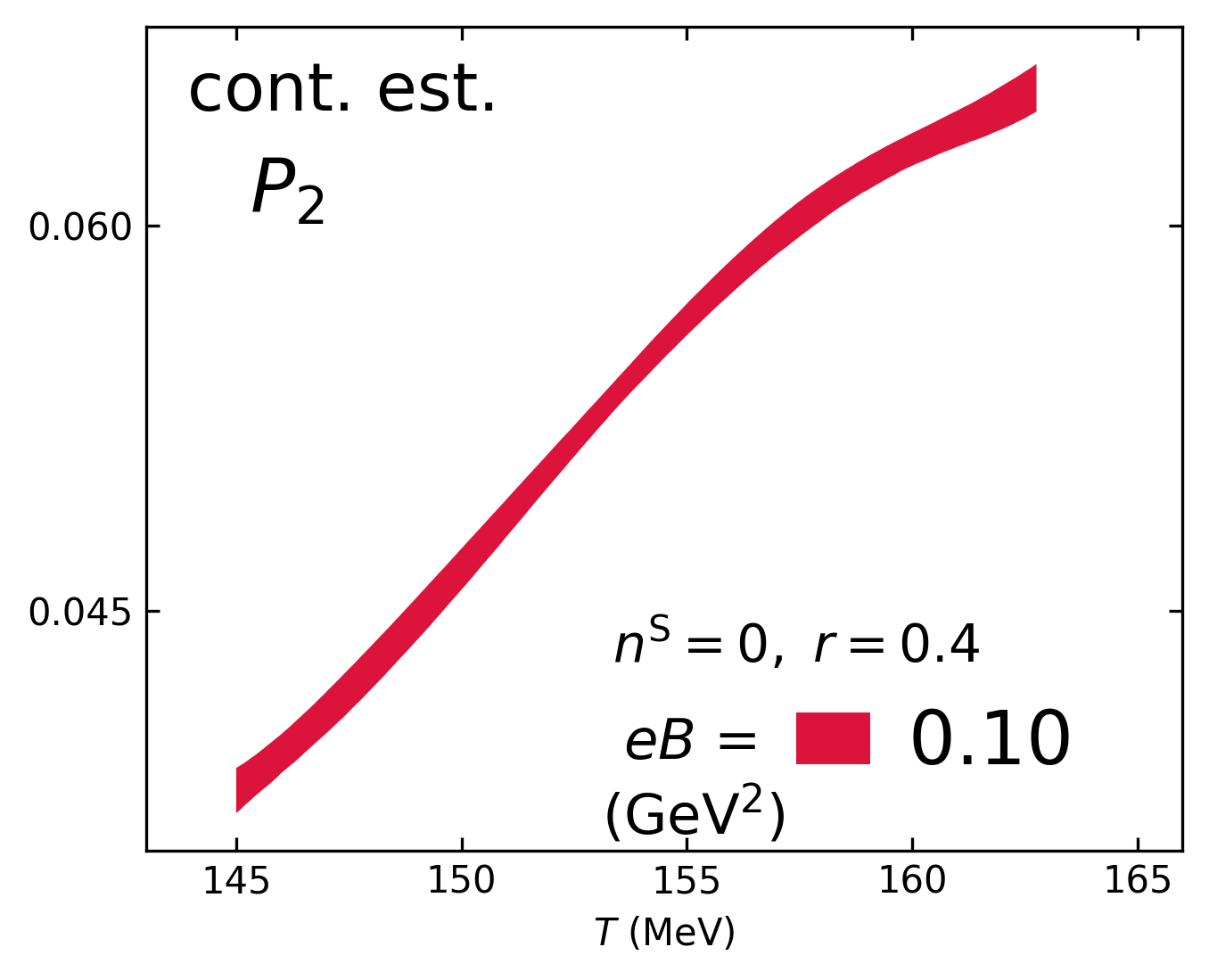}
\includegraphics[width=0.24\textwidth]{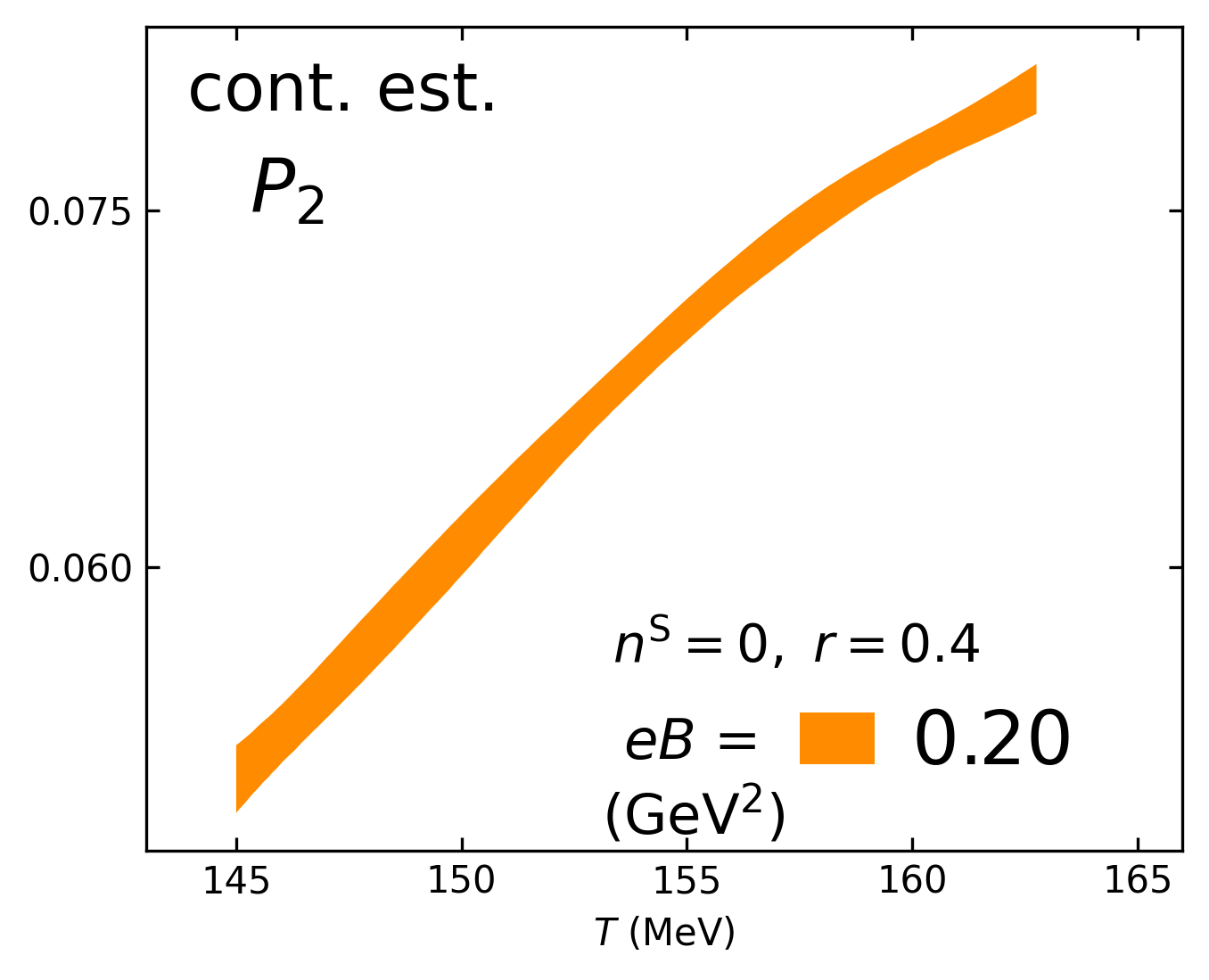}

\includegraphics[width=0.24\textwidth]{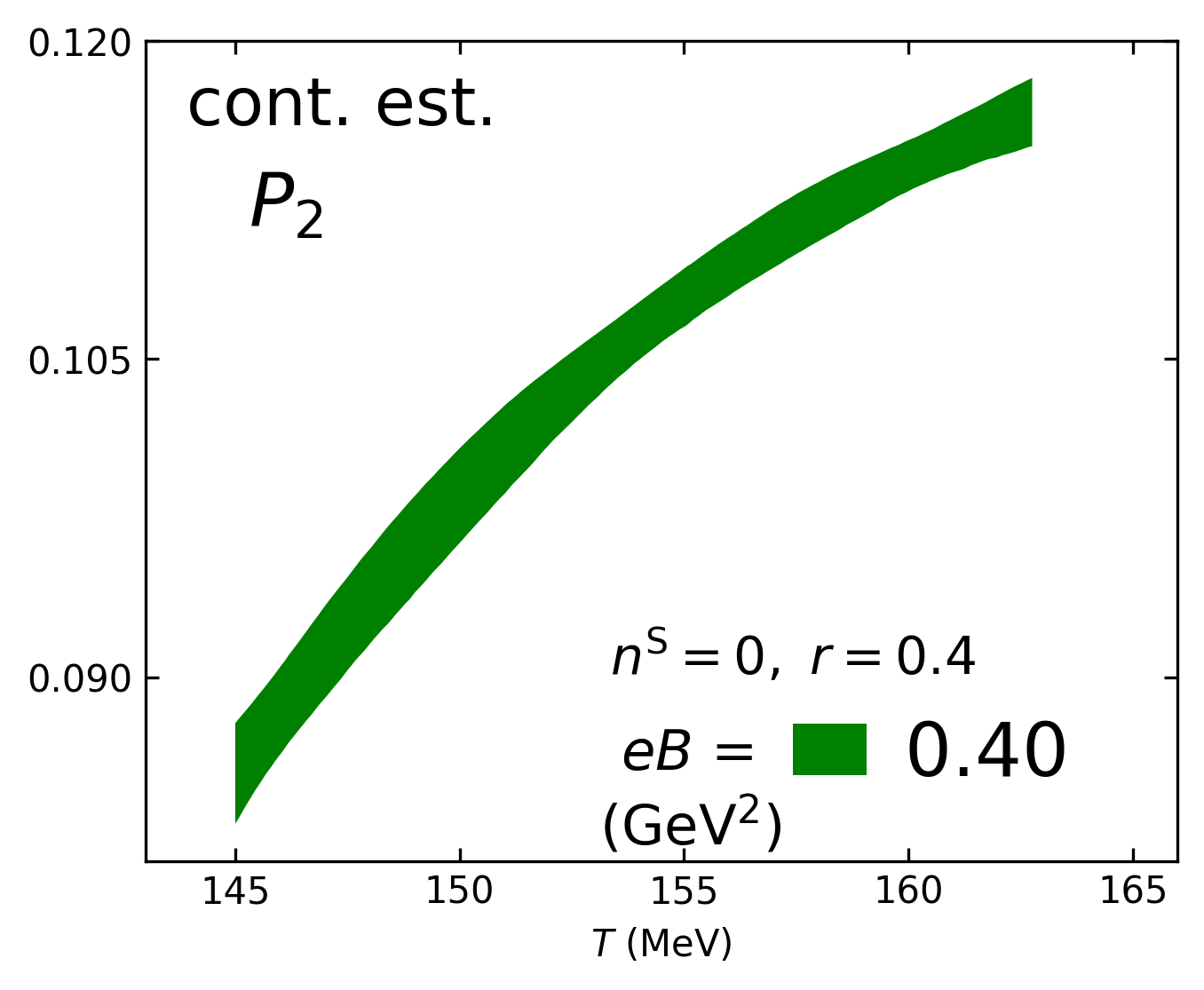}
\includegraphics[width=0.24\textwidth]{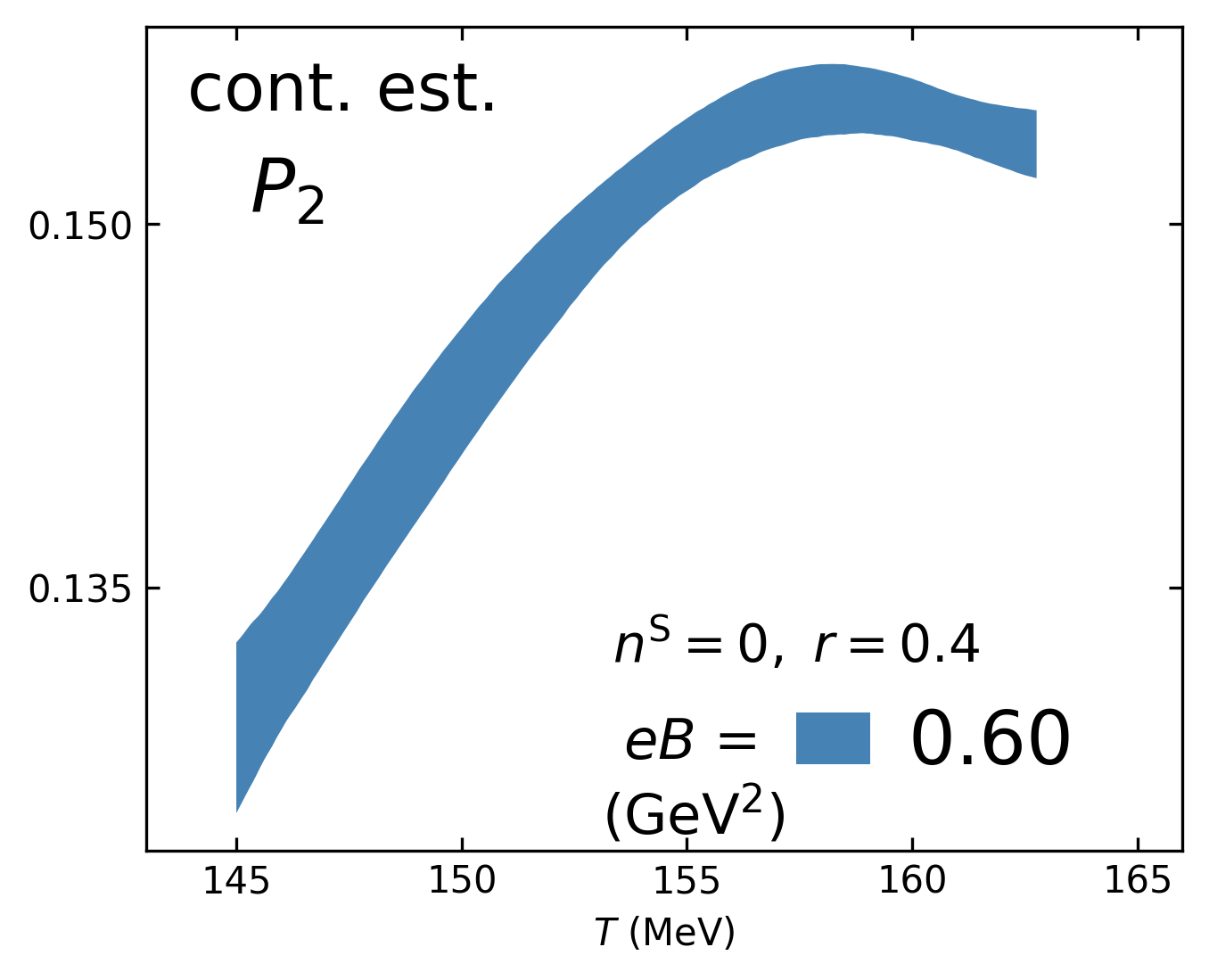}
\includegraphics[width=0.24\textwidth]{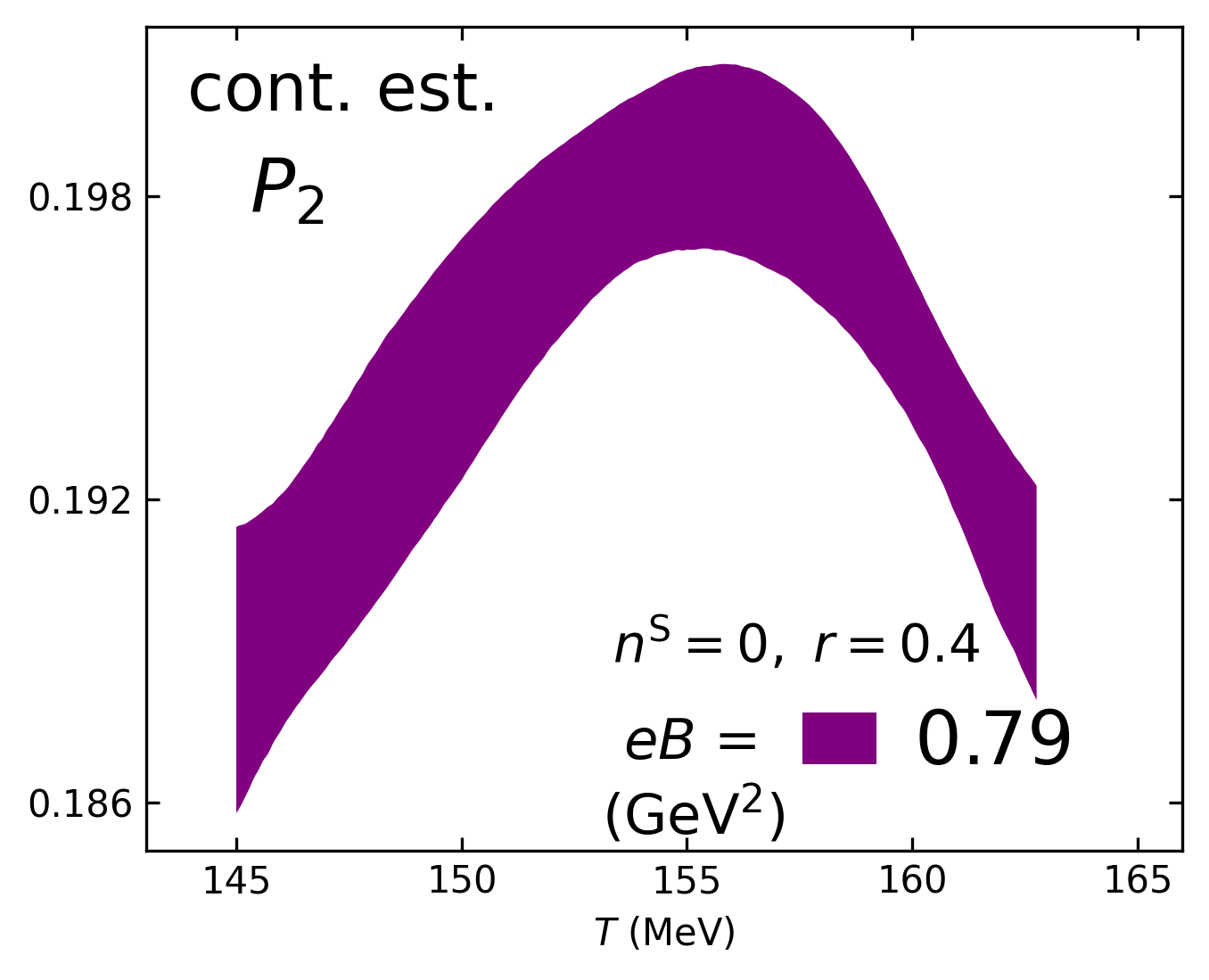}

\caption{Leading-order Taylor expansion pressure coefficient $P_2$ for strangeness-neutral systems $\hat{n}^{\rm S}=0$ with slight isospin asymmetry $r=0.4$. Figure is taken from Ref.~\cite{Ding:2025nyh}. }   
\label{fig:p2_Pb_Au_cont}
\end{figure}

Fig.~\ref{fig:p2_Pb_Au_cont} presents lattice QCD results for leading--order pressure coefficient
\begin{equation}
P_2 = \frac{1}{2!}\left( \chi^{{\rm B}}_{2}  + \chi^{{\rm Q}}_{2} q_1^2 +  \chi^{{\rm S}}_{2} s_1^2 \right)  + \chi^{{\rm B} {\rm Q}}_{11} q_1 + \chi^{{\rm B} {\rm S}}_{11}  s_1+ \chi^{{\rm Q} {\rm S}}_{11} q_1 s_1.
\end{equation}
At vanishing magnetic fields, it is well established that the pressure increases monotonically as the temperature increases due to the thermal agitations, with a pronounced rise near the QCD transition region. In the presence of magnetic fields, the behaviour of pressure is expected to be much more intricate due to the nontrivial interplay between thermal and magnetic field effects.

In the top-left panel, as the strength $eB$ grows, the pressure keeps increasing, stemming primarily from the fact that the degeneracy of Landau levels is directly proportional to the field strength. However, note that HRG interpretations are only applicable for the relatively weaker--$eB$ and low--$T$ regime. In the strong-$eB$ regime, around $eB\sim 0.6~{\rm GeV}^2$, we observe crossings among temperature bands which highlight a reordering of the temperature hierarchy, marking a qualitative departure from the monotonic hierarchy characteristic of the vanishing-- and weak--$eB$ regimes. 


In the right panel, magnetic field-induced reordering is reflected as the emergence of non-monotonic structures in the temperature profile. At even stronger magnetic fields, especially at $eB = 0.79~{\rm GeV}^2$, a mild peak structure becomes evident. This behaviour is further illustrated in the bottom panels for isolated magnetic strengths, clearly revealing the onset of non-monotonicity for $eB \gtrsim 0.6~{\rm GeV}^2$ and the formation of peak structure at $eB = 0.79~{\rm GeV}^2$. 
Lattice results in $T$--dependence indicate that both the inflection point of $P_2$ (typically associated with the pseudo-critical temperature $T_{pc}$) and the location of emerging non-monotonic peak structures systematically shift toward lower temperatures with increasing $eB$. This behaviour suggests that the transition in dominant degrees of freedom is driven to occur at reduced temperatures under stronger magnetic fields, commonly referred to as the $T_{pc}$-lowering effect, and is consistent with previous studies~\cite{Borsanyi:2023buy,Astrakhantsev:2024mat}. In the regime of extremely strong magnetic fields, lattice results progressively align with magnetized ideal gas predictions, represented by dashed colored lines, wherein the temperature hierarchy is ultimately reversed. Furthermore, unlike the ratio observables $q_1$ and $s_1$, there is no saturation approach with increasing $eB$, for pressure. Physically, this reflects the extensive nature of pressure, which increases as additional quantum states become accessible due to enhanced magnetic degeneracy.

\section{Summary}
In this work, we presented lattice QCD results of second-order fluctuations of conserved charges  in the presence of strong magnetic fields. Utilizing the state-of-the-art lattice QCD simulations with physical pion masses and performing continuum estimates, we explored magnetic field strengths extending up to unprecedented levels ($eB \simeq 0.8~\text{GeV}^2$). 
The baryon-electric charge correlation $\chi^{\rm BQ}_{11}$ stands out as a uniquely sensitive probe of magnetic fields in QCD matter, bridging theoretical predictions and experimental feasibility. Our work provides QCD benchmarks for interpreting current and future HIC data, particularly in disentangling the interplay of thermal and magnetic effects. These advances pave the way for a deeper exploration of QCD under extreme magnetic conditions, with implications for understanding the quark-gluon plasma and the role of magnetic fields in relativistic nuclear collisions.

Leading-order QCD EoS results demonstrate that strong magnetic fields qualitatively restructure the temperature dependence of leading-order bulk thermodynamic coefficients in strangeness-neutral QCD matter at finite baryon density. The emergence of temperature-band crossings, temperature-hierarchy reversals, and correlated patterns among different thermodynamic coefficients consistently reflect nontrivial interplay properties between thermal and magnetic effects.

\paragraph{Acknowledgements}
This work is supported partly by the National Natural Science Foundation of China under Grants No. 12293064, No. 12293060, and No. 12325508, as well as the National Key Research and Development Program of China under Contract No. 2022YFA1604900 and the Fundamental Research Funds for the Central Universities, Central China Normal University under Grants No. 30101250314 and No. 30106250152. The numerical simulations have been performed on the GPU cluster in the Nuclear Science Computing Center at Central China Normal University ($\mathrm{NSC}^{3}$) and Wuhan Supercomputing Center.

\bibliographystyle{elsarticle-num}
\bibliography{refs.bib}

\end{document}